\documentclass[twocolumn]{autart}
\usepackage{amsmath}
\usepackage{amsfonts}

\usepackage{algorithm,multirow,xcolor}
\usepackage{algorithmicx}
\usepackage{algpseudocode}

\usepackage{booktabs}
\usepackage{amssymb}% http://ctan.org/pkg/amssymb
\usepackage{pifont}%
\usepackage{graphicx}
\usepackage{subfigure}
\usepackage{verbatim}

\usepackage{dsfont}

\usepackage{mathrsfs}
\usepackage{times}
\usepackage{helvet}
\usepackage{courier}
\usepackage[hyphens]{url}
\usepackage{array}
\usepackage{color}
\usepackage{mathtools}

\usepackage{booktabs}       % professional-quality tables
\usepackage{nicefrac}       % compact symbols for 1/2, etc.

\usepackage[compress]{cite}
\newcommand{\define}{\triangleq}
\newtheorem{theorem}{Theorem}
\newtheorem{definition}{Definition}

\newtheorem{assumption}{Assumption}

\newtheorem{remark}{Remark}

\begin{document}
\begin{frontmatter}

\title{Reinforcement Learning Control of Constrained Dynamic Systems with Uniformly Ultimate Boundedness Stability Guarantee 
}

\thanks[footnoteinfo]{This paper was not presented at any IFAC 
meeting. Corresponding author W.~Pan. 
Email: \texttt{wei.pan@tudelft.nl}.}

\author[lxz-a]{Minghao Han},  
\author[wp-b]{Yuan Tian}, 
\author[lxz-a]{Lixian Zhang}, 
\author[jw-c]{Jun Wang}, 
\author[wp-b]{Wei Pan}
\vspace{-0.05in}
\address[lxz-a]{Department of Control Science and Engineering, Harbin Institute of Technology, China.}
\address[wp-b]{Department of Cognitive Robotics, Delft University of Technology, Netherlands.}
\address[jw-c]{Department of Computer Science, University College London, UK.}

%%%%%%%%%%%%%%%%%%%%%%%%%%%%%%%%%%%%%%%%%%%%%%%%%%%%%%%%%%%%%%%%%%%%%%%%%%%%%%%%
\begin{abstract}
Reinforcement learning (RL) is promising for complicated stochastic nonlinear control problems. Without using a mathematical model, an optimal controller can be learned from data evaluated by certain performance criteria through trial-and-error. However, the data-based learning approach is notorious for not guaranteeing stability, which is the most fundamental property for any control system. In this paper, the classic Lyapunov's method is explored to analyze the uniformly ultimate boundedness stability (UUB) solely based on data without using a mathematical model. It is further shown how RL with UUB guarantee can be applied to control dynamic systems with safety constraints. Based on the theoretical results, both off-policy and on-policy learning algorithms are proposed respectively. As a result, optimal controllers can be learned to guarantee UUB of the closed-loop system both at convergence and during learning. The proposed algorithms are evaluated on a series of robotic continuous control tasks with safety constraints. In comparison with the existing RL algorithms, the proposed method can achieve superior performance in terms of maintaining safety. As a qualitative evaluation of stability, our method shows impressive resilience even in the presence of external disturbances.

\end{abstract}
\end{frontmatter}

%%%%%%%%%%%%%%%%%%%%%%%%%%%%%%%%%%%%%%%%%%%%%%%%%%%%%%%%%%%%%%%%%%%%%%%%%%%%%%%%
\section{Introduction}

The recent progress in reinforcement learning (RL) \cite{sutton1992reinforcement} has produced many interesting and impressive results in control problems and proves to be effective in finding optimal controllers for nonlinear stochastic systems modeled by Markov decision process (MDP), for which the traditional control methods are hardly applicable. However, the learning methods are notorious for not guaranteeing stability. \textcolor{black}{Given a control system, stability is one of the most important properties, because an unstable system is typically useless and potentially dangerous.} This presents a major bottleneck for the broad control engineering applications. Stability analysis has a long history in control engineering, in which Lyapunov's method plays a central role  \cite{slotine1991applied,vidyasagar2002nonlinear,sastry2013nonlinear}. However, the classical control methods rely on the full or partial knowledge of the system dynamics to design controllers and are largely limited to systems with simple dynamics. Thus, it is a natural move to combine RL with control theory to develop learning control methods with a stability guarantee \cite{bucsoniu2018reinforcement,han2020actor}.

\begin{figure}[tb]
    \centering
    \includegraphics[width=0.98\columnwidth]{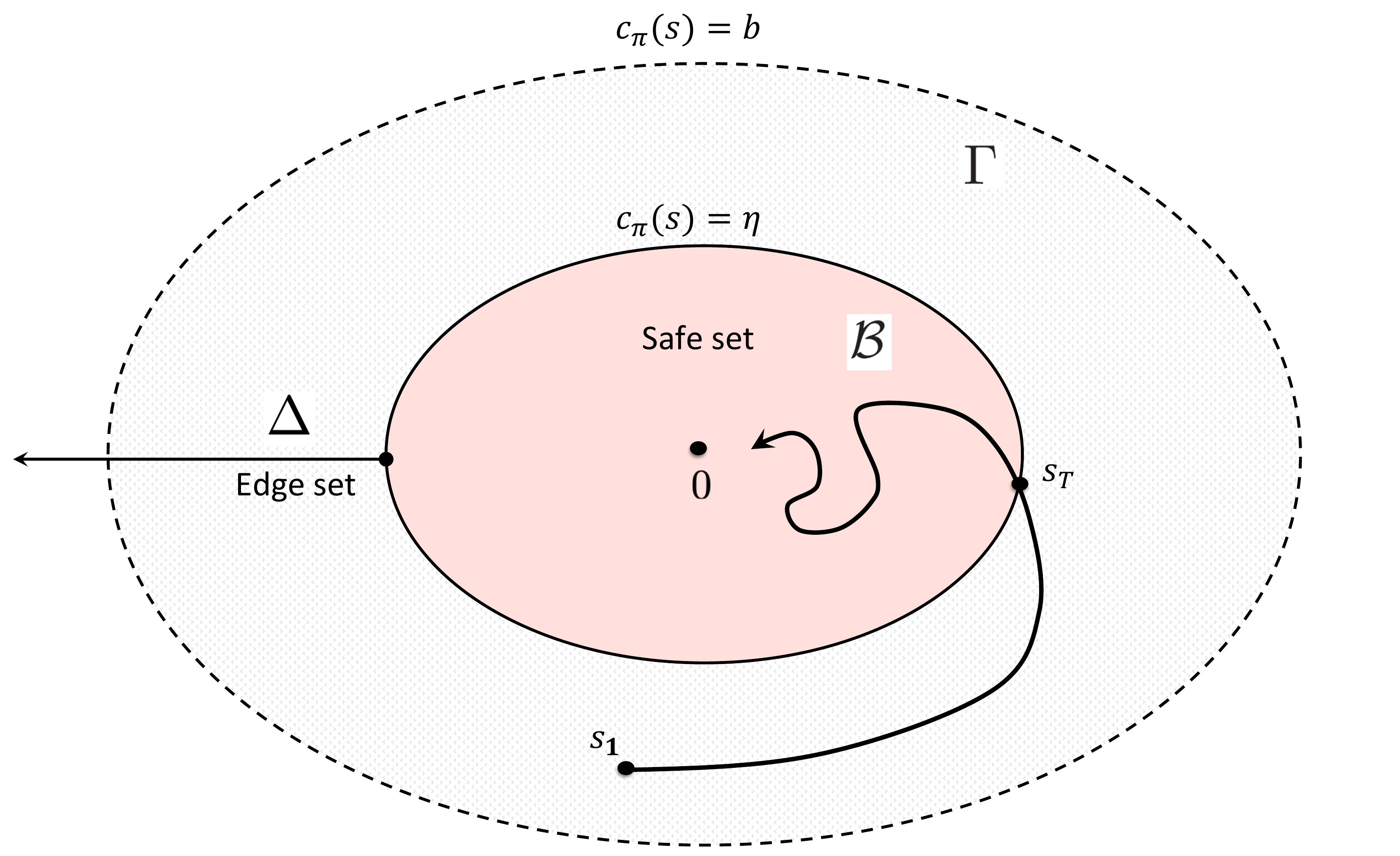}
    \caption{Conceptual illustration of UUB in the state space and the relation among the starting set $\Gamma$, the edge set $\Delta$ and the inner set $\mathcal{B}$. For trajectories starting from the set $\Gamma: \{s|\Vert s\Vert\leq b\}$, the state will eventually enter and stay inside the inner set $\mathcal{B}:\{s|\Vert s\Vert\leq \eta\}$ after $T$ time steps.}
    \label{fig:state space}
\end{figure}

Among various definitions of stability, uniformly ultimate boundedness stability (UUB) has been extensively studied for dynamic systems \cite{corless1981continuous,jain2017uniformly}. UUB generally says that the trajectories will enter the neighborhood of the equilibrium within finite time and never escape from this set thereafter (see the trajectory in Fig.~\ref{fig:state space}). Intuitively, this property is consistent with the requirement of many control tasks with constraints on the states, where the states are required to stay in a certain region which is safe.  Thus in this paper, as an application scenario, the controller with UUB guarantee is learned to solve control tasks with safety constraints. In terms of learning a controller, UUB is well known in the context of adaptive dynamic programming (ADP) on (1) UUB of the states or tracking error in uncertain systems \cite{shih2007near,yang2016guaranteed,mu2016data} and (2) UUB of the estimation error of critic function and controller \cite{shih2007near,wang2018event,luo2019event}. However, in ADP, the model structure, not necessarily the model parameters, should be known a prior. And the input structure of the nonlinear system is often assumed to be affine. However, the UUB analysis for the general class of nonlinear stochastic systems by solely using data has not been addressed and remains as an open problem~\cite{bucsoniu2018reinforcement,gorges2017relations}.

The control tasks with safety constraints on the state have been extensively studied in model predictive control (MPC) literature \cite{mayne2000constrained,mayne2001control} and the results are applied in various industrial processes \cite{garcia1989model,scattolini2009architectures}. In the context of RL, control problems with safety constraints are also well studied. In \cite{achiam2017constrained}, the authors proposed a safety constrained policy optimization (CPO) approach based on the trust region method, which guarantees the constraint satisfaction with a safe initial policy, but it is restricted to the on-policy algorithms and suffers from the low sample efficiency.  In \cite{chow2018lyapunov}, a Lyapunov-based approach for solving constrained control tasks is proposed with a novel way of constructing the Lyapunov function through linear programming. In \cite{chow2019lyapunov}, the above result is further generalized to continuous control tasks. However, the above results can only guarantee that the cumulative sum of a designed constraint function \textcolor{black}{being kept} under a threshold. Moreover, none of these results provides stability guarantees of any kind. Since the property of attraction is missing, simply satisfying the safety constraints does not imply stability, and the agent may easily violate constraints in the presence of slight disturbances. 

\textcolor{black}{To apply machine learning algorithms to control constrained dynamic systems is advancing recently. In~\cite{berkenkamp2017safe}, a model-based RL method is proposed to deal with Lipschitz continuous deterministic nonlinear systems. Nevertheless, safety is ensured by validating the stability condition on discretized points in the subset of state space with the help of a learned model, limiting its application to rather simple and low-dimensional systems. The combination between RL with control barrier functions (CBF) have raised many attentions in recent years \cite{cheng2019end,choi2020reinforcement}. The general idea is to incorporate the RL controller with a model-based controller using CBFs. \cite{cheng2019end} exploits the nominal model to design a CBF-based controller to ensure safety, then the unknown dynamic is learned using the Gaussian process and the RL controller further improves the return performance. In \cite{choi2020reinforcement}, based on the controller designed using the nominal model, RL is exploited to solve the safe control problems in control affine nonlinear systems under model uncertainty. Another common way of solving constrained control tasks is to incorporate MPC with online model-learning techniques, such as  \cite{ostafew2016robust,thananjeyan2020safety,zanon2020safe}. \cite{saunders2018trial} proposes an RL framework that can exploit expert knowledge to safely improve control performance without violating safety constraints. Different from these approaches, in this paper, neither the nominal model nor expert knowledge is needed to learn a safe controller.}

In this paper, a novel data-based UUB theorem without using a mathematical model is proposed. Based on the theoretical result, an off-policy based on an actor-critic algorithm and a policy optimization algorithm are developed respectively to learn controllers with the UUB guarantee.
The contributions of this paper can be summarized as follows: 
\begin{itemize}
    \item  A novel and principled method is proposed to construct Lyapunov functions based on data to analyze the closed-loop stability of stochastic nonlinear systems characterized by MDP.
    \item The classical definition of UUB is generalized to deal with control tasks with safety constraints on the states.
    \item Practical algorithms are designed to search \textcolor{black}{for} the optimal safe policy with UUB guarantee while safety is guaranteed both during learning and exploitation.
\end{itemize}
In a series of high-dimensional continuous control tasks with safety constraints such as locomotion for legged robots and \textcolor{black}{manipulators}, as well as a quadrotor, the proposed algorithms outperform the existing (safe) RL algorithms \cite{achiam2017constrained,chow2019lyapunov} in terms of both performance and safety. Besides, it is empirically shown that the controller with the UUB guarantee is more capable of dealing with perturbations and disturbances in comparison with those without such guarantees.

The remainder of this paper is organized as follows: the preliminaries and problem formulation are introduced in Section~\ref{sec:preliminaries}; the main theoretical result is presented in Section~\ref{sec:stability}; an off-policy algorithm and a policy optimization algorithm with the UUB guarantee are described in Section~\ref{sec:algorithm}; the experiments that validate the proposed algorithms are presented in Section~\ref{sec:experiments};  Section~\ref{sec:conclusion} concludes this work.

\section{Preliminaries}\label{sec:preliminaries}

In RL, a dynamical system is often characterized by a Markov decision process (MDP) in which the next state only depends on the current state and action. \textcolor{black}{In MDPs}, $s_t\in\mathcal{S}\subseteq \mathbb{R}^n$ is the state vector at time $t$, $\mathcal{S}$ denotes the state space. The agent then takes an action $a_t \in\mathcal{A}\subseteq \mathbb{R}^m$ according to a stochastic policy/controller\footnote{We use controller throughout the paper.} $\pi(a_t|s_t)$, resulting in the next state $s_{t+1}$. The transition of the state is dominated by the transition probability density function $p(s_{t+1}|s_t,a_t)$, which denotes the probability density of the next state $s_{t+1}$. 
In MDP, a reward function $r(s_t,a_t)$ is used to measure \textcolor{black}{the immediate performance of} a state-action pair $(s_t,a_t)$. The goal is to find $\pi$ which can maximize the objective function/return $J(\pi) \triangleq  \sum_{t=1}^{\infty} \mathbb{E}_{s_t,a_t} \gamma^t r(s_t,a_t)$. Additionally, for control problems with safety constraints, a continuous non-negative constraint function $c(s_t,a_t)$ is introduced to measure how safe a state-action pair is. The state-action pair can be viewed as safe if $c(s_t,a_t)$ is lower than a designed threshold.

\begin{figure}[tb]
    \centering
    \includegraphics[width=0.9\columnwidth]{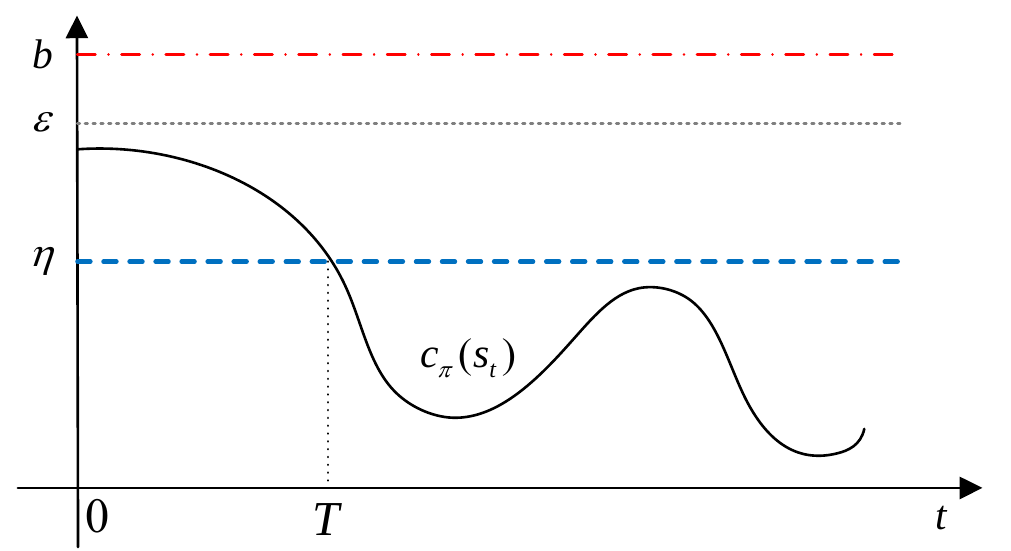}
    \caption{An illustration of the concept of UUB in time domain.
    }
    \label{fig:timeview}
\end{figure}

Next, the \textcolor{black}{notation} of this paper will be introduced in two parts. First, the concept and definition of UUB studied in this paper are introduced. Then the constrained control problems or so-called constrained Markov decision process (CMDP) are described as a particular application scene of UUB.

\subsection{Uniformly Ultimate Boundedness (UUB) Stability}

First, the classical definition of UUB stability is given as follows.
\begin{definition}\label{definition:classical uub}
\cite{thowsen1983uniform}
A system is said to be uniformly ultimately bounded with ultimate bound $\eta$, if there exist positive constants $b$, $\eta$, $\forall\epsilon<b$, there exists $T(\epsilon, \eta)$, such that $\Vert s_{t_0}\Vert<\epsilon \implies \Vert s_t\Vert <\eta, \forall t> t_0+T$. If this holds for arbitrary large $\epsilon$, then it is globally uniformly ultimately bounded.
\end{definition}

The classical definition of UUB generally says that for trajectories starting from a point where the norm of \textcolor{black}{the} state less than $b$, the state will eventually enter and stay inside the set where $\Vert s\Vert\leq \eta$ after $T$ time steps.

However, the above definition is of limited use for the general class of control tasks with safety constraints, \textcolor{black}{where the constraint functions $c(s_t,a_t)$ are not necessarily the norm of the state, i.e., $\Vert s\Vert$}. For example, safety-critical applications like autonomous vehicles may use the distance from the working area or central lane as the safety constraint; altitude control of a drone may require that sinusoid of \textcolor{black}{an} angle to be less than a certain threshold. Therefore, the classical definition of UUB is extended to the more general case as follows.

\begin{definition}
\label{def:uub}
A system is said to be uniformly ultimately bounded with respect to $c_\pi(\cdot)$, if there exist positive constants $\eta$, $b$, $\forall\epsilon<b$, there exists $T(\epsilon, \eta)$, such that $c_\pi(s_{1})\leq \epsilon \implies c_\pi(s_{t})\leq \eta, \forall t\geq T$.
\end{definition}
where $c_\pi(s) \triangleq \mathbb{E}_{a\sim\pi }c(s,a)$ denotes the constraint function function under the controller $\pi$. In the rest of this paper, UUB refers to the property defined above.

The difference between Definition~\ref{definition:classical uub} and Definition~\ref{def:uub} is merely on the substitution of the norm of states with a constraint function. The general idea of UUB in Definition~\ref{def:uub} is demonstrated in Fig.~\ref{fig:state space} and \ref{fig:timeview} in the state space and time domain respectively. With the proper choice of $c(s,a)$, UUB in terms of the constraint function implies recoverability from danger within finite time. For example, a vehicle will recover to the road within finite time if it is accidentally disturbed and run into a risky area; a motor can recover to normal status within finite time if the kinetic energy or torque accidentally exceeds a dangerous threshold.

\subsection{Control of Constrained Dynamic System}\label{sec:problem}
In this section, the control problems with safety constraints will be formulated. It will be further shown how safety constraints can be ensured as a consequence of the UUB guarantee.

The safety constraints are measured by a continuous non-negative constraint function $c(s_t,a_t)$. In the safety constrained control tasks, the objective is to find a controller $\pi$ which not only maximizes $J(\pi)$ but also satisfy the expectation of the safety constraint $\mathbb{E}_{s_t}c_\pi(s_t)\leq d,\forall t\in[1,\infty)$, i.e.,
\begin{equation}
    \max_\pi J(\pi) \text{  s.t. } \mathbb{E}_{s_t}c_\pi(s_t)\leq d,\forall t\in[1,\infty).
    \label{eq:safety constrained control problem}
\end{equation}
\textcolor{black}{An} MDP with such safety constraints is called constrained Markov decision process (CMDP) \cite{altman1999constrained}.

If the system is UUB in terms of the constraint function with an ultimate bound $\eta<d$, the value of the constraint function is guaranteed to converge and stay under $\eta$ after $T$ time steps. To further ensure safety during the $T$ time steps, it is also needed to ensure that the expectation of the constraint function to be lower than $d$. Later it will be shown that this is an inherent property of the system being UUB. Thus solving the control problem \eqref{eq:safety constrained control problem} is equivalent to finding an optimal controller that ensures the closed-loop system is UUB.

Before proceeding, some notations are introduced. The action-value function $Q_\pi(s,a)\triangleq  \sum_{t=1}^{\infty} \mathbb{E}_{s_t,a_t} [\gamma^t r(s_t,a_t)|s_{1}=s,a_1=a]$ denotes the subsequent return under the controller $\pi$ after taking action $a$ at the state $s$. The value function with respect to constraint function $V^c_\pi(s)\triangleq  \sum_{t=1}^{\infty} \mathbb{E}_{s_t,a_t} [\gamma^t c(s_t,a_t)|s_{1}=s]$ denotes the discounted sum of constraint function starting from the state $s$ under the controller $\pi$. $\rho (s_1)$ denotes the probability density function of starting states, which is a continuous function and takes positive values on the starting set $\Gamma\triangleq\{s|c_\pi(s)\leq b\}$. The closed-loop transition probability is denoted as $p_\pi(s'|s)\define\int_\mathcal{A}\pi(a|s)p(s'|s,a)\mathrm{d}a$. Also note that the closed-loop state distribution at a certain instant $t$ as $p(s|\rho, \pi, t)$, which can be defined iteratively: $p(s'|\rho, \pi, t+1) =\int_\mathcal{S} p_\pi(s'|s)p(s|\rho, \pi, t)\mathrm{d}s, \forall t\in\mathbb{Z}_{[1,\infty)}$ and $p(s|\rho, \pi, 1) =\rho(s)$.

Two important sets are exploited in this paper, as shown in Fig.~\ref{fig:state space}. First, the edge set $\Delta \triangleq\left\{ s|c_\pi(s) \geq \eta\right\}$ is \textcolor{black}{composed of states that are unsafe}. Conversely, the inner set $\mathcal{B}\triangleq\{s|c_\pi(s)<\eta\}$ is the set of absolutely safe states and $\mathcal{B}\cup \Delta = \mathcal{S}$.

\section{Theoretical Results}\label{sec:stability}

In this section, the main assumptions and a novel data-based UUB theorem for stochastic systems are proposed. The UUB theorem enables \textcolor{black}{analyzing} the UUB of the system through a data-based manner, which further enables policy learning in the next section.

First, we made the following assumption:
\begin{assumption}\label{assumption:cost}
The constraint function $c(s,a)$ is non-negative and there exists an integrable function $g(s,a)$ such that $c(s,a)\leq g(s,a)$, $\forall (s,a)\in \mathcal{S}\times \mathcal{A}$.
\end{assumption}

This assumption easily holds with the typical choices of constraint function in practice, such as kinetic energy, altitude, etc. It is also assumed that the Markov chain induced by policy $\pi$ is ergodic with a unique stationary distribution $q_\pi$, $$q_\pi(s)=\lim_{t\rightarrow\infty}p(s|\rho,\pi,t),$$  which is a common assumption for the optimal control of a Markov decision process \cite{sutton2009convergent, korda2015td, bhandari2018finite, zou2019finite}.

Our approach utilizes the Lyapunov function to prove the stability condition. Lyapunov's method has long been used in control theory for stability analysis and controller design \cite{boukas2000robust}, but mostly exploited along with a known model of a dynamic system, whether deterministic or probabilistic \cite{thowsen1983uniform,corless1981continuous,huang2011uniformly}.

In this paper, instead of using a model, we will present the following theorem as a sufficient condition for UUB solely based on samples.
\begin{theorem}\label{theorem:uub}
If there exists a function $L(s):\mathcal{S}\rightarrow \mathbb{R_+}$ and positive constants $\alpha_{1}$, $\alpha_{2}$, $\alpha _{3}$, $\eta$, such that
\begin{equation}
\alpha _{1}c_\pi(s) \leq L(s)\leq \alpha _{2}c_\pi(s),\forall s\in \mathcal{S}  \label{theorem:uub-1}
\end{equation}%
and
\begin{align}
\begin{aligned}
&\mathbb{E}_{s\sim \mu_N}\left( \mathbb{E}_{s^{\prime }\sim P_{\pi
}}L(s^{\prime })\mathds{1}_\Delta(s^\prime )-L(s)\mathds{1}_\Delta(s)\right) \\
< &-\alpha _{3}\mathbb{E}_{s\sim \mu_N}c_\pi(s)\mathds{1}_\Delta(s) 
\end{aligned}
\label{theorem:uub-2}
\end{align}%
where $\mu_N(s)$ denotes the average distribution of $s$ over the finite $N$ time steps,
\begin{equation*}
    \mu_N(s)\doteq\frac{1}{N}\sum_{t=1}^N p(s|\rho, \pi,t)
\end{equation*}
$N$ is the \textcolor{black}{maximum instant that the probability of being in the edge set is greater than zero, $N=\max \{t: \mathbb{P}(s\in \Delta|\rho, \pi, t)>0\}$; $N=\infty$ if for any $\delta$ there exists an instant $t>\delta$ such that $\mathbb{P}(s\in \Delta|\rho, \pi, t)>0$}. $\mathds{1}_\Delta(s)$ denotes the function
\begin{equation*}
   \mathds{1}_\Delta(s) =\left\{
\begin{array}{cc}
1 & s\in\Delta \\
0 & s\notin\Delta%
\end{array}%
\right.
\end{equation*}
where $\Delta =\left\{ s|c_\pi(s) \geq \eta\right\}$.

Then one has the following hold: i) the system is uniformly ultimately bounded with ultimate bound $\eta$; ii) the expectation $\mathbb{E}_{p(s|\rho, \pi, t)}c_\pi(s)$ is bounded during the $N$ time steps, $\mathbb{E}_{p(s|\rho, \pi, t)}c_\pi(s)<\frac{\alpha_2b}{\alpha_3}+ \eta,\forall t\in[1,N]$.
\end{theorem}

\textbf{Proof. }
\textcolor{black}{The proof can be divided into two steps. First, we will prove that $N$ is finite based on the conditions and assumptions, then prove the boundedness on the expectation of $c_\pi$ during the $N$ steps.}
To show this, we will assume that $N$ is infinity and prove by contradiction.

In that case, the finite-horizon sampling distribution $\mu_N(s)$ turns into the infinite-horizon sampling distribution $$\mu(s)= \lim_{N\rightarrow \infty}\mu_N(s)=\lim_{N\rightarrow \infty}\frac{1}{N}\sum_{t=1}^N p(s|\rho, \pi,t)$$
The existence of $\mu(s)$ is guaranteed by the existence of $q_\pi(s)$. Since the sequence $\{p(s|\rho,\pi,t), t\in\mathbb{Z}_+\}$ converges to $q_\pi(s)$ as $t$ approaches $\infty$, then by the Abelian theorem, the sequence $\{\frac{1}{T}\sum_{t=1}^T p(s|\rho,\pi,t), T\in\mathbb{Z}_+\}$ also converges and $\mu(s) = q_\pi(s)$.
Then one naturally has that the sequence $\{\mu_N(s)L(s)\mathds{1}_\Delta(s), T\in\mathbb{Z}_+\}$ converges pointwise to $q_\pi(s)L(s)\mathds{1}_\Delta(s)$. 

Let $g_\pi(s)$ denote $\mathbb{E}_{a\sim\pi}g(s,a)$.
According to Assumption~\ref{assumption:cost} and \eqref{theorem:uub-1}, $L(s)\leq \alpha_2c_\pi(s)\leq\alpha_2g_\pi(s)$, which follows that 
\begin{align*}
\mu_N(s)L(s)\mathds{1}_\Delta(s)&\leq\alpha_2\mu_N(s)g_\pi(s)\mathds{1}_\Delta(s)   
\end{align*}
According to the Lebesgue's Dominated convergence theorem\cite{royden1968real}, if a sequence ${f_n(s)}$ converges point-wise to a function $f$ and is dominated by some integrable function $g$ in the sense that,
\begin{equation*}
\vert f_n(s) \vert \leq g(s), \forall s\in \mathcal{S},\forall n
\end{equation*}
then one has
\begin{equation*}
\lim_{n\rightarrow\infty}\int_\mathcal{S}f_n(s)\mathrm{d}s= \int_\mathcal{S}\lim_{n\rightarrow\infty}f_n(s)\mathrm{d}s
\end{equation*}

Applying this theorem to the first term in the left-hand-side of \eqref{theorem:uub-2}
\begin{align*}
    &\mathbb{E}_{s\sim \mu}\mathbb{E}_{s^{\prime}\sim p_{\pi}}L(s^\prime)\mathds{1}_\Delta(s^\prime)\\
    =&\int_\mathcal{S}\lim_{N\rightarrow\infty}\frac{1}{N}\sum_{t=1}^N p(s|\rho, \pi,t)(\int_\mathcal{S} p_{\pi}(s^{\prime}|s)L(s^{\prime})\mathds{1}_\Delta(s^\prime)\mathrm{d}s^{\prime})\mathrm{d}s\\
    = &\lim_{N\rightarrow\infty}\frac{1}{N}\sum_{t=1}^N \int_\mathcal{S}L(s^{\prime})\mathds{1}_\Delta(s^\prime) \int_\mathcal{S}p_{\pi}(s^{\prime}|s)p(s|\rho, \pi,t) \mathrm{d}s\mathrm{d}s^{\prime}\\
    =&\lim_{N\rightarrow\infty}\frac{1}{N}\sum_{t=2}^{N+1} \mathbb{E}_{ p(s|\rho, \pi,t)}L(s)\mathds{1}_\Delta(s)
\end{align*}
Similarly, $\mathbb{E}_{s\sim \mu}L(s)=\lim_{N\rightarrow\infty}\frac{1}{N}\sum_{t=1}^{N} \mathbb{E}_{ p(s|\rho, \pi,t)}L(s)$.
It follows that on the left-hand-side of (\ref{theorem:uub-2}),
{\small
\begin{align*}
     &\mathbb{E}_{s\sim \mu}(\mathbb{E}_{s^{\prime}\sim p_{\pi}}L(s^\prime)\mathds{1}_\Delta(s^\prime)-L(s)\mathds{1}_\Delta(s))\\
     =&\lim_{N\rightarrow\infty}\frac{1}{N}(\sum_{t=2}^{N+1} \mathbb{E}_{ p(s|\rho, \pi,t)}L(s)\mathds{1}_\Delta(s)-\sum_{t=1}^{N} \mathbb{E}_{ p(s|\rho, \pi,t)}L(s)\mathds{1}_\Delta(s))\\
     =&\lim_{N\rightarrow\infty}\frac{1}{N}(\mathbb{E}_{p(s|\rho, \pi,N+1)}L(s)\mathds{1}_\Delta(s)- \mathbb{E}_{ \rho(s)}L(s)\mathds{1}_\Delta(s))
\end{align*}
}

Since $\mathbb{E}_{ \rho(s)}L(s)\mathds{1}_\Delta(s)$ is finite, thus the limitation value $\lim_{N\rightarrow\infty}\frac{1}{N}(\mathbb{E}_{ \rho(s)}L(s)\mathds{1}_\Delta(s)))=0$. The above equation equals to 
\begin{align*}
    &\lim_{N\rightarrow\infty}\frac{1}{N}\mathbb{E}_{ p(s|\rho, \pi,N+1)}L(s)\mathds{1}_\Delta(s)\\
    \geq&\lim_{N\rightarrow\infty}\frac{\alpha_1}{N}\mathbb{E}_{ p(s|\rho, \pi,N+1)}c_\pi(s)\mathds{1}_\Delta(s)
\end{align*}
Note that $c_\pi(s)\mathds{1}_\Delta(s)$ is greater than $\eta$ if $s\in\Delta$ and equals to zero if $s\notin\Delta$, which can be summarized as $c_\pi(s)\mathds{1}_\Delta(s)\geq\eta\mathds{1}_\Delta(s)$. Thus
\begin{align}
\begin{aligned}
    &\lim_{N\rightarrow\infty}\frac{1}{N}\mathbb{E}_{ p(s|\rho, \pi,N+1)}L(s)\mathds{1}_\Delta(s)\\
    \geq&\lim_{N\rightarrow\infty}\frac{\alpha_1\eta}{N}\mathbb{E}_{ p(s|\rho, \pi,N+1)}\mathds{1}_\Delta(s)\\
    =&\lim_{N\rightarrow\infty}\frac{\alpha_1\eta}{N}\mathbb{P}(s\in\Delta|\rho, \pi,N+1)\\
    =&0
\end{aligned}\label{proof:1}
\end{align}

Now let's look into the right-hand-side of \eqref{theorem:uub-2}. Since $\mu(s)=q_\pi(s)$, the right-hand-side of \eqref{theorem:uub-2} equals to 
\begin{align*}
    &-\alpha_{3}\mathbb{E}_{s\sim q_\pi}c_\pi(s)\mathds{1}_\Delta(s)\\
    \leq&-\alpha_{3}\mathbb{E}_{s\sim q_\pi}\eta\mathds{1}_\Delta(s)\\
    =&-\alpha_{3}\eta\lim_{t\rightarrow\infty}\mathbb{P}(s\in\Delta|\rho, \pi,t)
\end{align*}
Combining the above inequality with \eqref{theorem:uub-2} and \eqref{proof:1}, one has that $\lim_{t\rightarrow\infty}\mathbb{P}(s\in\Delta|\rho, \pi,t)<0$, which is contradictory with the fact that $\mathbb{P}(s\in\Delta|\rho, \pi,t)$ is non-negative. Thus there exist a finite $N$ such that $\mathbb{P}(s\in\Delta|\rho, \pi,t) =0$ for all $t>N$, which concludes the proof of UUB.

Additionally, it will be shown that the expectation of $c_\pi(s)$ is bounded by a finite value during the $N$ time steps.
As $N$ is a finite value, \eqref{theorem:uub-2} implies that
\begin{align*}
    &\mathbb{E}_{p(s|\rho, \pi,N+1)}L(s)\mathds{1}_\Delta(s)- \mathbb{E}_{ \rho(s)}L(s)\mathds{1}_\Delta(s)\\
    <& -\alpha_3\sum_{t=1}^{N} \mathbb{E}_{ p(s|\rho, \pi,t)}c_\pi(s)\mathds{1}_\Delta(s)
\end{align*}
Then for any instant $n\in[1,N]$, one has the following hold
\begin{align*}
 &\mathbb{E}_{ p(s|\rho, \pi,n)}c_\pi(s)\mathds{1}_\Delta(s)\\
    \leq&\frac{\alpha_2}{\alpha_3}\mathbb{E}_{\rho(s)}c_\pi(s) -\sum_{t=n+1}^{N} \mathbb{E}_{ p(s|\rho, \pi,t)}c_\pi(s)\mathds{1}_\Delta(s)\\
    &-\sum_{t=1}^{n-1} \mathbb{E}_{ p(s|\rho, \pi,t)}c_\pi(s)\mathds{1}_\Delta(s)
\end{align*}
Note that the expectation of $c_\pi(s)$ at instant $n$ equals to 
\begin{align*}
    &\int_\mathcal{S}p(s|\rho, \pi,n)c_\pi(s)\mathrm{d}s \\
    = &\int_\Delta p(s|\rho, \pi,n)c_\pi(s)\mathrm{d}s + \int_\mathcal{B}p(s|\rho, \pi,n)c_\pi(s)\mathrm{d}s\\
    =&\mathbb{E}_{ p(s|\rho, \pi,n)}c_\pi(s)\mathds{1}_\Delta(s) + \int_\mathcal{B}p(s|\rho, \pi,n)c_\pi(s)\mathrm{d}s
\end{align*}
Then the bound of the expectation of $c_\pi(s)$ at instant $n$ is derived as follows,
\begin{align*}
    &\mathbb{E}_{ p(s|\rho, \pi,t)}c_\pi(s)\\
    \leq&\frac{\alpha_2}{\alpha_3}\mathbb{E}_{\rho(s)}c_\pi(s) -\sum_{t=n+1}^{N} \mathbb{E}_{ p(s|\rho, \pi,t)}c_\pi(s)\mathds{1}_\Delta(s)\\
    &-\sum_{t=1}^{n-1} \mathbb{E}_{ p(s|\rho, \pi,t)}c_\pi(s)\mathds{1}_\Delta(s)+ \int_\mathcal{B}p(s|\rho, \pi,n)c_\pi(s)\mathrm{d}s\\
    \leq&\frac{\alpha_2b}{\alpha_3} -\eta\sum_{t=n+1}^{N}\mathbb{P}(s\in\Delta|\rho, \pi,t)\\
    &- \eta\sum_{t=1}^{n-1}\mathbb{P}(s\in\Delta|\rho, \pi,t) + \eta \mathbb{P}(s\in\mathcal{B}|\rho, \pi,n)\\
    <&\frac{\alpha_2b}{\alpha_3}+ \eta
\end{align*}
which concludes the proof.
$\hfill \square $

Some discussion and explanations are needed for the above theorem. First, \eqref{theorem:uub-1} confines the property that the Lyapunov function needs to satisfy. \eqref{theorem:uub-2} is the data-based energy decreasing condition, of which the evaluation requires sampling data according to sampling distribution $\mu_N(s)$. Although $\mu_N(s)$ is defined on the state space $\mathcal{S}$, the indication function $\mathds{1}_\Delta(s)$ only takes the non-zero value on the edge set $\Delta$. Thus \eqref{theorem:uub-2} requires the Lyapunov value to be decreasing on large in the edge set $\Delta$ and eventually entering the inner set $\mathcal{B}$.

\begin{remark}
While results on UUB of various systems are well-known \cite{thowsen1983uniform,corless1981continuous,huang2011uniformly}, however, both of these results require the full knowledge of the dynamic model of the system. On the contrary, the proposed UUB theorem enables a data-based approach to analyze the stability of the system, i.e. collecting lots of state transition pairs and evaluate the value of \eqref{theorem:uub-2} through the Monte-Carlo method. In the data-based stability analysis, the system can be a complete black-box, as long as its dynamic satisfies the Markov property.
\end{remark}

Some connections are to be drawn between safety constrained control problems and the proposed UUB theorem. If the system is UUB with ultimate bond $\frac{\alpha_2b}{\alpha_3}+ \eta<d$, then it is guaranteed that the system satisfies the safety constraint in \eqref{eq:safety constrained control problem}. These conditions can be satisfied by choosing the hyperparameters $\alpha_2$, $\alpha_3$, and $\eta$. \eqref{theorem:uub-2} is the condition that requires training of the control policy, which will be discussed in detail in the following section.

\section{Reinforcement Learning Algorithms with UUB Stability Guarantee}\label{sec:algorithm}
In this section, combined with the theoretical result in Theorem~\ref{theorem:uub}, both an off-policy and an on-policy RL algorithm are proposed respectively. 
First, based on the result in Theorem~\ref{theorem:uub}, an actor-critic RL algorithm called Lyapunov-based soft actor-critic (LSAC) is given, where two critic functions are used. The first is the standard critic function $Q(s,a)$ in the actor-critic RL algorithm, which is used to evaluate the performance in terms of the cumulative return. The other critic function is introduced to evaluate the UUB condition \eqref{theorem:uub-2}. We call the second critic function as Lyapunov critic function $L_c(s,a)$.
Then a trust-region policy optimization algorithm, Lyapunov-based constrained policy optimization (LCPO) is developed. LCPO ensures that at each update step the UUB condition is satisfied and increases the cumulative return monotonically so that the safety during training \textcolor{black}{can} also be guaranteed. \textcolor{black}{In both LSAC and LCPO, a Lyapunov function is firstly specified to direct learn the controller and Lyapunov function. The controller is then updated to ensure that the energy decreasing UUB condition~\eqref{theorem:uub-2} holds for the learned Lyapunov function.}

\subsection{Learning a Lyapunov Critic Function}
\textcolor{black}{In Theorem~\ref{theorem:uub}, the Lyapunov function $L(s)$ is essential in the stability analysis. However, it is not directly applicable in an existing actor-critic learning framework since the gradient of $L$ with respect to the controller $\pi$ is unavailable. To enable the actor-critic learning, the Lyapunov critic function $L_c(s,a)$ is introduced to prove the stability theorem therefore make sure the learned controller $\pi$ can guarantee the stability of the closed-loop system. $L_c$ depends on both the state $s$ and the action $a$\footnote{It should be noted that the Lyapunov critic function $L_c$ is \emph{not} a proper Lyapunov function since it also depends on action $a$.}. $L_c$ satisfies $L(s) = \mathbb{E}_{a\sim \pi} L_c(s,a)$, such that it can be exploited by judging the value of \eqref{theorem:uub-2}.} In this paper, $L_c$ is constructed by using a fully connected deep neural network (DNN) parameterized by $\phi$. A ReLU activation function is used in the output layer of the DNN to ensure positive output.

From a theoretical point of view, some functions, such as the norm of state and value function, naturally satisfy the basic requirement of Lyapunov function \eqref{theorem:uub-1}. These functions are referred to as Lyapunov candidate functions. 
However, Lyapunov candidate functions are conceptual functions without any parameterization. Since their gradient with respect to the controller is not tractable, they are not directly applicable in an actor-critic learning process. In the proposed framework, the Lyapunov candidate acts as a supervision signal during the training of $L_c$.
$L_c$ is updated to approximate the target function $L_{\text{target}}$ related to the chosen Lyapunov candidate, minimizing the following objective function simply using least square algorithm,
\begin{equation}\label{eq:critic objective}
    J(\phi) = \mathbb{E}_{ \mathcal{D}}\left[\frac{1}{2}(L_c(s,a)-L_{\text{target}}(s,a))^2\right]
\end{equation}
where $\mathcal{D}=\{(s,a,s',r, c)\}$ is the set of observed transition tuple under the controller $\pi$.

The choice of the Lyapunov candidate plays an important role in learning a controller. In control theory, the sum of quadratic polynomials, e.g., $L(s)=s^\top Ps$ where $P$ is a positive definite matrix, is often used. Such Lyapunov functions can be efficiently discovered by using semi-definite programming (SDP) solvers with certain limited conservatism for control tasks. In the context of RL \cite{chow2018lyapunov,berkenkamp2017safe}, the value function $V^c_\pi$ is proved to be a valid Lyapunov candidate. In the meantime, the constraint function $c_\pi$ is also a valid Lyapunov candidate due to its nonnegativity. The value function and constraint function are chosen to be the Lyapunov candidate in this paper while other potential choices are left for future study.

With the value function chosen to be the Lyapunov candidate, the target function $L_{\text{target}}$ in \eqref{eq:critic objective} is 
\begin{equation}
L_{\text{target}}(s,a) = c(s,a) + \max_{a'}\gamma L'_c(s', a')
\end{equation}
where $L'_c$ is the network that has the same structure as $L_c$, but parameterized by a different set $\phi^{\prime}$. These parameters of the neural network is updated through exponentially moving average controlled by a hyperparameter $\tau\in \mathbb{R}_{(0,1)}$, $\phi^{\prime}_{k+1}\leftarrow \tau\phi_{k} + (1-\tau)\phi^{\prime}_{k}$, as typically used in actor-critic algorithms~\cite{lillicrap2015continuous}.
\textcolor{black}{It should be noted that when the constraint function is chosen to be the Lyapunov candidate, the target function $L_{\text{target}}$ is much simpler by having $L_{\text{target}}(s,a) = c(s,a)$, and the $L'_c$ network and moving average update are not needed.}

\textcolor{black}{
\begin{remark}
This remark will collectively summarize the Lyapunov terms used in the section above for clarity.
i) $L$ refers to the Lyapunov function. 
ii) Lyapunov candidates are functions that potentially can be used as a Lyapunov function, such as value function and $\Vert s \Vert$. 
iii) The Lyapunov critic function $L_c$ is a function dependent on the state and action, and satisfies $L = \mathbb{E}_{a\sim \pi} L_c(s,a)$. 
iv) The target function $L_{target}$ refers to the supervision signal used in the training of $L_c$, which takes different forms when different Lyapunov candidates are chosen.
v) $L_c'$ is a network that shares the same structure with $L_c$, but only the parameters are updated by applying moving average to the parameter of $L_c$. This is a typical trick used in the RL literature, designed to improve the stability of the learning process.
\end{remark}}

\textcolor{black}{
\begin{remark}
In this paper, the Lyapunov functions are parameterized using neural networks.
It is also possible to choose a parameterization form that can be updated using SDP, such as a quadratic of $s$ and $a$. However, such a parameterization may result in a large approximation error when the constraint function involves multiple types of nonlinearities that can't be approximated using a single polynomial, e.g. the constraint functions with
non-differentiable nonlinearities that will be used Section 5. Alternatively, neural networks are powerful function approximators that can theoretically approximate any nonlinear functions in desired precision, thus we exploit neural networks to show the general applicability of the proposed method, and leave other parameterizations for future studies.
\end{remark}}

\subsection{Lyapunov-based Safe Off-Policy RL Algorithm}

A novel off-policy RL algorithm based on the actor-critic algorithm, i.e., Lyapunov Safe Actor-Critic (LSAC), is proposed to learn the controller $\pi$ that can maximize the return while guaranteeing the UUB for the closed-loop system.

The controller $\pi_\theta$ is parameterized by a DNN $f_\theta(s,\epsilon)$ depending on $s$ and a Gaussian noise $\epsilon$. The goal is to learn $\theta$ that can maximize $J(\pi)$ in \eqref{eq:safety constrained control problem}, while satisfying the UUB condition \eqref{theorem:uub-2} simultaneously.
In this paper, we build the actor-critic algorithm based on the maximum entropy framework \cite{haarnoja2018soft-b}, which can enhance the exploration of the controller during learning and has been proven to substantially improve the robustness of the learned controller \textcolor{black}{\cite{haarnoja2018soft,ma2020distributional}}. The learning problem is summarized as follows, 
\begin{align}
    \max_\theta \text{ }&\mathbb{E}_{\mathcal{D}}Q_{\pi_\theta}(s,a)\\
    \text{s.t. }& \eqref{theorem:uub-2}\\ &-\mathbb{E}_{\mathcal{D}}\log\pi_\theta(a|s)\geq\mathcal{H}_t\label{eq:minimum entropy}
\end{align}
where \eqref{eq:minimum entropy} sets the entropy of the controller to be larger than a designed threshold $\mathcal{H}_t$. By exploiting the Lagrange method, solving the above constrained optimization problem is equivalent to minimizing the following objective function,
\begin{equation}
\begin{aligned}
J(\pi) =& \mathbb{E}_{  \mathcal{D}}\left[-Q(s,f_\theta(s,\epsilon)) + \beta \log\pi_\theta(f_\theta(s,\epsilon)|s) \right] \\
& + \lambda\mathbb{E}_{\mathcal{D}_\Delta}[ L_c(s^{\prime },f_\theta(s^{\prime},\epsilon))\mathds{1}_\Delta(s^\prime )\\
& -(L_c(s,a) -\alpha_3c)\mathds{1}_\Delta(s)]
\label{LSAC_policy_cost}
\end{aligned}
\end{equation}
where $\beta$ and $\lambda$ are positive Lagrangian multipliers. Both the values of $\beta$ and $\lambda$ are adjusted through the gradient descent/ascent method. $\mathcal{D}_\Delta$ denotes the transition pairs collected from the sampling distribution $\mu_N(s)$.

In our implementation, the double Q-learning technique \cite{van2016deep} is exploited, where two Q-functions $\{Q_1,Q_2\}$ are parameterized by neural networks with parameters $\nu_1$ $\nu_2$. The Q-function with the lower value is exploited in the learning process \cite{fujimoto2018addressing}, which is useful in mitigating performance degradation caused by the bias in the value estimation. Taking these techniques into consideration, the gradient concerning $\theta$ is obtained as
\begin{equation}
\begin{aligned}
    \nabla_\theta J(\pi)=&\mathbb{E}_{\mathcal{D}}[-\min_iQ_i(s,a)\nabla_\theta f_\theta(s',\epsilon) \\
    &+\beta\nabla_\theta  \log(\pi_\theta(a|s)) + \beta\nabla_a \log\pi_\theta(a|s)]\\
    &+\lambda\mathbb{E}_{\mathcal{D}_\Delta}[\nabla_{a'}L_c(s',a')\nabla_\theta f_\theta(s',\epsilon)\mathds{1}_\Delta(s^\prime )]
\end{aligned}\label{eq:policy gradient}
\end{equation}
The gradient is composed of two parts: (1) the gradient estimated by the Q-function and the entropy of the controller based on the samples from replay buffer $\mathcal{D}$, and (2) the gradient estimated by the Lyapunov critic based on the samples from the edge buffer $\mathcal{D}_\Delta$. \eqref{eq:policy gradient} is the basis of the actor-critic algorithm and enables the update of controller with observed transition pairs.

In the actor-critic algorithm, the Q-function is updated by using gradient descent to minimize the following objective function
\begin{equation*}
J(Q_i) = \mathbb{E}_{\mathcal{D}}\frac{1}{2}\left[r+\gamma Q'_i(s',f_\theta(s',\epsilon))-Q_i(s,a)\right]^2, \ i\in \{1,2\}
\end{equation*}
where $Q'_i$ is the target network that has the same structure with $Q_i$ and parameterized by $\nu'_i$ but updated through moving average.

Finally, the value of Lagrange multipliers $\beta$ and $\lambda$ are adjusted by gradient ascent to maximize the following objectives, respectively,
\begin{gather*}
J(\beta) = \beta\mathbb{E}_{\mathcal{D}}  [\log\pi_\theta( a|s)+\mathcal{H}_t]\label{eq:temperature update}\\
J(\lambda) =\lambda\mathbb{E}_{\mathcal{D}_\Delta}[ L_c(s^{\prime },f_\theta(s^{\prime},\epsilon))\mathds{1}_\Delta(s^\prime ) -(L_c(s,a) -\alpha_3c)\mathds{1}_\Delta(s)]
\end{gather*}

The pseudocode of LSAC is summarized in Algorithm~\ref{algo:LSAC}.
\begin{algorithm}[htb]
   \caption{Lyapunov-based Safe Actor-Critic Algorithm (LSAC)}
   \label{algo:LSAC}
\begin{algorithmic}
    \State Set iteration index $i \leftarrow 0$ 
    and learning rate $\delta$ \\
    \Repeat 
    \State Sample $s_1$ according to $\rho$ \\
    \For{each time step}
    \State Sample $a_t$ from $\pi(a_t|s_t)$ and step forward \\
    \State Observe and store $(s_t,a_t,r_t,c_{t},s_{t+1})$ in $\mathcal{D}$\\
   \EndFor
   \State Record the largest instant $N$ at which $s_N\in\Delta$
   \State Store all tuples $(s_t,a_t,r_t,c_{t},s_{t+1}), t\leq N$ in $\mathcal{D}_\Delta$
   \For{each update step}
   \State Sample mini-batches of transitions from $\mathcal{D}$ and $\mathcal{D}_\Delta$ and update parameters with gradients,
        \begin{align*}
            \theta &\leftarrow \theta -  \delta \nabla_\theta J(\pi)\notag \\
            \phi &\leftarrow \phi -  \delta \nabla_\phi J(L_c) \notag \\
            \nu_i &\leftarrow \nu_i -  \delta \nabla_{\nu_i} J(Q_i) \notag \\
            \lambda &\leftarrow \lambda +  \delta \nabla_\lambda J(\lambda) \notag \\
            \beta &\leftarrow \beta +  \delta \nabla_{\beta} J(\beta) \notag 
        \end{align*}
   \EndFor
   \State $i\leftarrow i+1$
   \Until{\eqref{theorem:uub-2} is satisfied and $i$ exceeds a designed threshold.}

\end{algorithmic}
\end{algorithm}

\subsection{Lyapunov-based On-Policy RL Algorithm}
The off-policy algorithms can exploit the data collected under a different controller and update the controller much more frequently than the on-policy algorithms, and thus is more favorable in terms of data efficiency and convergence speed. However, safety can be hardly guaranteed during the training process of off-policy algorithms \cite{chow2019lyapunov}. When the controller is trained online in the real-world and data are collected directly from the physical systems, safety needs to be guaranteed even during training. To this end, an on-policy algorithm called Lyapunov Constrained Policy Optimization (LCPO) is proposed for these safety-critical scenarios.

In comparison with LSAC \eqref{eq:policy gradient}, instead of approximating the policy gradient using the approximated critics, LCPO is a trust-region style method built upon the linearization of constraint and objective functions locally around the current parameter $\theta$. In trust-region methods, a local constrained optimization problem is solved at each update step and the parameter updates a small step towards improving the objective while satisfying constraints. \textcolor{black}{This determines that the optimization of policy needs more samples than the actor-critic methods to correctly approximate the constrained optimization problem,} as well as more computation time in the line-search at each update \cite{achiam2017constrained}. However, on the other hand, this also guarantees the approximate monotonic improvement of the performance and safety during training with an initial safe policy \cite{moldovan2012safe,amodei2016concrete}. Following such a procedure, it is possible to directly train a controller on the real system with safety being assured. The update of the policy parameter $\theta$ at the $k_{\text{th}}$ iteration can be formulated as follows
\begin{align}
\theta_{k+1} &=\arg \max_{\theta }\mathbb{E}_{\substack{s\sim \mathcal{D} \\ a\sim \pi}} Q_{\pi _{k}}\left(s,a\right) \label{eq:objective}\\
\text{s.t. }& \mathbb{E}_{\substack{s\sim \mathcal{D}_\Delta \\ a\sim \pi }}\left[ L(s^{\prime
})\mathds{1}_\Delta(s^{\prime
})-\left(L(s)-\alpha_3 c\right) \mathds{1}_\Delta(s)\right]\leq 0 \label{eq:safety constraint}\\
& \mathbb{E}_{\mathcal{D}}D_{\text{KL}}\left( \pi_\theta |\pi_k \right)\leq \delta
\label{eq:local policy search constraint}
\end{align}
Here, $\pi_k$ denotes the policy parameterized by $\theta_k$ at $k_\text{th}$ update. In trust-region method, there is a local policy search constraint \eqref{eq:local policy search constraint} to prevent the policy from taking unreasonably large update steps, and ensure that post-update policy stays in the neighborhood of the previous policy specified by $\delta$. Here, $D_{KL}(p|q)$ denotes the KL-divergence between two distributions $p$ and $q$, $D_{KL}(p|q)\doteq\mathbb{E}_p\log (p/q)$. The KL-divergence is a measure of the difference between two distributions and is commonly used in trust-region methods \cite{schulman2015trust,achiam2017constrained}.
At each update step, the above constrained optimization problem is solved analytically. Since the search of policy is constrained around the previous policy $\pi_k$ by \eqref{eq:local policy search constraint}, it is possible to linearize the objective function \eqref{eq:objective} and the safety constraint \eqref{eq:safety constraint} around $\pi_k$ and approximate the local policy search constraint \eqref{eq:local policy search constraint} using second-order expansion \cite{achiam2017constrained}.
The approximated optimization problem is as follows,
\begin{align}
\begin{aligned}
\theta _{k+1} =&\arg \max_{\theta }g_Q^{\top}\left( \theta -\theta _{k}\right)
\\
\text{s.t. }&g_L^{\top}\left( \theta -\theta _{k}\right)+h \leq 0 \\
&\frac{1}{2}\left( \theta -\theta _{k}\right) ^{\top}H\left( \theta -\theta _{k}\right)
\leq \delta
\end{aligned}
\label{eq:approximated}
\end{align}%
where $g_Q$ and $g_L$ are gradients of the objective function and the safety constraint function with respect to $\theta$ at $\theta_k$, $h$ is the value of the safety constraint function at $\theta_k$ and $H$ is the Hessian of the KL-divergence. Note that the Fisher information matrix $H$ is guaranteed to be positive semi-definite, thus the above optimization problem is convex and its dual is as follows,
\begin{align}
\max_{\lambda, \beta \geq 0} \frac{1}{2\beta}(g_Q^T H^{-1}g_Q - 2 \lambda\mathcal{Z} + \lambda^2 \mathcal{N})+ \lambda h - \frac{\beta\delta}{2} \label{eq:dual problem}
\end{align}%
where $\mathcal{Z} = g_Q^T H^{-1}g_L$ and $\mathcal{N}=g_L^T H^{-1}g_L$. Suppose that the original problem is feasible and $\lambda^*$ and $\beta^*$ are the solutions to \eqref{eq:dual problem}, \textcolor{black}{then the optimal solution to the primal problem \eqref{eq:approximated} is given by} 
\begin{equation}
    \theta_{k+1} = \theta_k + \frac{1}{\beta^*}H^{-1}(g_Q - \lambda^* g_L)
    \label{eq:optimal solution}
\end{equation}
If the optimization problem \eqref{eq:approximated} is not feasible, then a recovery update step is needed. For safety constrained tasks, it is important that the policy $\pi$ recovers to a set of safe policies as soon as possible. In the meantime, \eqref{eq:local policy search constraint} needs to be satisfied as this is the basis of local approximate optimization. The recovery step is equivalent to solving the following optimization problem,
\begin{equation}
\begin{aligned}
\theta _{k+1} =&\arg \min_{\theta }g_L^{\top}\left( \theta -\theta _{k}\right)+h \\
\text{s.t. }&\frac{1}{2}\left( \theta -\theta _{k}\right) ^{\top}H\left( \theta -\theta _{k}\right)
\leq \delta
\end{aligned}
\label{eq:recovery optimization problem}
\end{equation}
The optimal solution to the above recovery optimization problem is 
\begin{equation}
    \theta^* = \theta_k -\sqrt{\frac{2\delta}{g_L^{\top}H^-1g_L}}H^-1g_L
    \label{eq:recovery step}
\end{equation}
In LCPO, the Lyapunov function $L(s)$ is also a DNN parameterized by $\phi$. The Lyapunov critic function $L_c$ is not needed since LCPO does not involve critic-actor updates. Meanwhile, the Lyapunov function candidates are still valid following a similar approximation procedure as LSAC. The pseudo-code of LCPO can be found in Algorithm~\ref{algo:LCPO}.
\begin{figure*}[htb]
    \centering
	\includegraphics[width=0.99\textwidth]{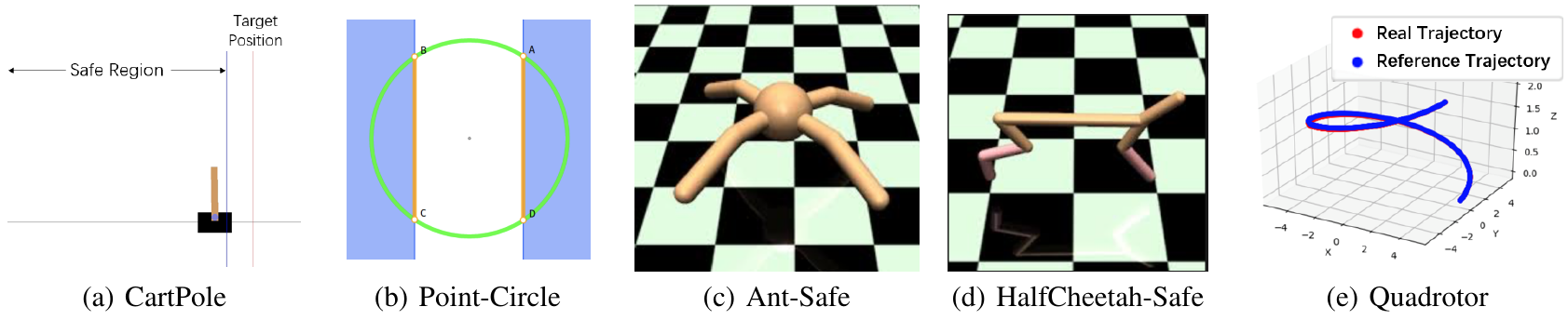}
    \caption{Demonstration of environments screenshots.}
    \label{fig:experiments_setup}
\end{figure*}

\begin{algorithm}[htb]
   \caption{Lyapunov-based Constrained Policy Optimization (LCPO)}
   \label{algo:LCPO}
\begin{algorithmic}
    \For{$i=1,2,\dots$}
%   \FOR{each iteration}
   
   \State Sample $s_1$ according to $\rho$
   \For{each time step}
   \State Sample $a_t$ from $\pi(a_t|s_t)$ and step forward
   \State Observe and store $(s_t,a_t,r_t,c_{t},s_{t+1})$ in $\mathcal{D}$
   \EndFor
   \State Record the largest instant $N$ at which $s_N\in\Delta$
   \State Store all tuples $(s_t,a_t,r_t,c_{t},s_{t+1}), t\leq N$ in $\mathcal{D}_\Delta$
   \State Estimate $g_Q$, $g_L$, $h$, $H$ with $\mathcal{D}$ and $\mathcal{D}_\Delta$
   \If{\eqref{eq:approximated} is feasible}
   \State Calculate the optimal $\lambda^*$ and $\beta^*$
   \State Calculate the proposal $\theta^*$ with \eqref{eq:optimal solution}
   \Else
   \State Calculate $\theta^*$ with \eqref{eq:recovery step}
   \EndIf
   \State Update $\theta_{k+1}$ by backtracking linesearch to satisfy the sample estimate \eqref{eq:safety constraint}
   \State Clear $\mathcal{D}$ and $\mathcal{D}_\Delta$
   \EndFor
\end{algorithmic}
\end{algorithm}
\begin{remark}
The comparison between LCPO and LSAC in terms of data efficiency is to be made. As shown in the pseudo-code Algorithm~\ref{algo:LSAC}, LSAC updates multiple steps after a trajectory has been sampled. It is even possible to update at every step after observing a new state-action pair, though this is not adopted in LSAC. In comparison, LCPO only proceeds one step after each iteration of observing the trajectory. This is due to the nature of on-policy algorithms: after one update of $\theta$, the collected data become off-policy data, i.e. the data generated by a different controller, and can not be used by the on-policy algorithms anymore. Thus, at the end of each iteration, LCPO needs to empty the set of transition pairs $\mathcal{D}$ and $\mathcal{D}_\Delta$. On the contrary, LSAC repeatedly makes use of the data collected by different controllers. As a result, LSAC possesses better data efficiency than LCPO.
\end{remark}

\begin{remark}
Another distinguishing difference between LCPO and LSAC is the safety guarantee during training, \textcolor{black}{which is a key consequence between off-and on-policy algorithms}. LCPO is implemented based on the trust-region optimization, solving an approximated optimization problem locally and assuring that every post-update policy is safe. \textcolor{black}{If the initial policy is safe, the safety will be ensured during learning. Otherwise, LCPO will try to find a safe policy first using the recovery update in Eq.\eqref{eq:recovery step}.} In comparison, LSAC does not hold this property but only can assure safety at the end of the training. From a practical point of view, LSAC is suitable for training a safe policy in a virtual environment and then deployed to a real system; LCPO is directly applicable for online training. Furthermore, a potential choice is to combine the strengths of LSAC and LCPO: use LSAC to learn an initial policy in the virtual environment then transfer to LCPO for further online training. 
\end{remark}

\subsection{\textcolor{black}{Further Discussion} }
\textcolor{black}{
Before proceeding to show the effectiveness of LSAC and LCPO for various environments, we would like to further discuss the possible adverse effects of sample-based approximation and UUB analysis.} 

\textcolor{black}{
First, in both LSAC and LCPO, the safety constraint is evaluated using a sample estimate of the inequality \eqref{theorem:uub-2}, which unavoidably introduces estimation error unless the number of samples is infinite. Therefore, a possible research direction is to establish a quantitative relation between the reliability of the safety guarantee and the number of samples. Furthermore, stronger safety constraints can be introduced by considering the estimation error. }

\textcolor{black}{
Second, as both LSAC and LCPO rely on the sample-based gradient approximation in controller training, the learning algorithms may take update steps in undesirable directions temporarily due to some approximation errors (such as reducing the return or violating the safety constraints). Nevertheless, this effect can be modified by choosing reasonable hyperparameters such as learning rate and batch size such that the undesirable update does not affect the convergence of the learning process.}

\textcolor{black}{Finally, different from the classical model-based controller design methods, the proposed method does not need a dynamic model to design the controller. Instead, it is model-free which means only the data from the trial and error will be used to learn the controller until a satisfactory one can be found. This process is undoubtedly time-consuming and hardly applicable to a system in the real world. Thus in practice, it is favorable to train the controller virtually first, then transfer and fine-tune the controller in the real world \cite{yu2019sim,tan2018sim,harrison2020adapt}. }

\section{Experiments}\label{sec:experiments}
In our experiments, we would like to address the following questions:
\begin{itemize}

\item How does our approach perform compared with other existing safe RL algorithms for CMDP tasks?

\item Can the algorithms converge with different parameter initializations and learn safe control policies in the presence of function approximation error?

\item Does the policy with UUB guarantee ensure the system to recover to the inner set under perturbation and disturbance?

\item Under what circumstances the Lyapunov-based safe RL algorithms may fail?

\end{itemize}

\begin{figure*}[htb]
\centering
\includegraphics[width=1.0\textwidth]{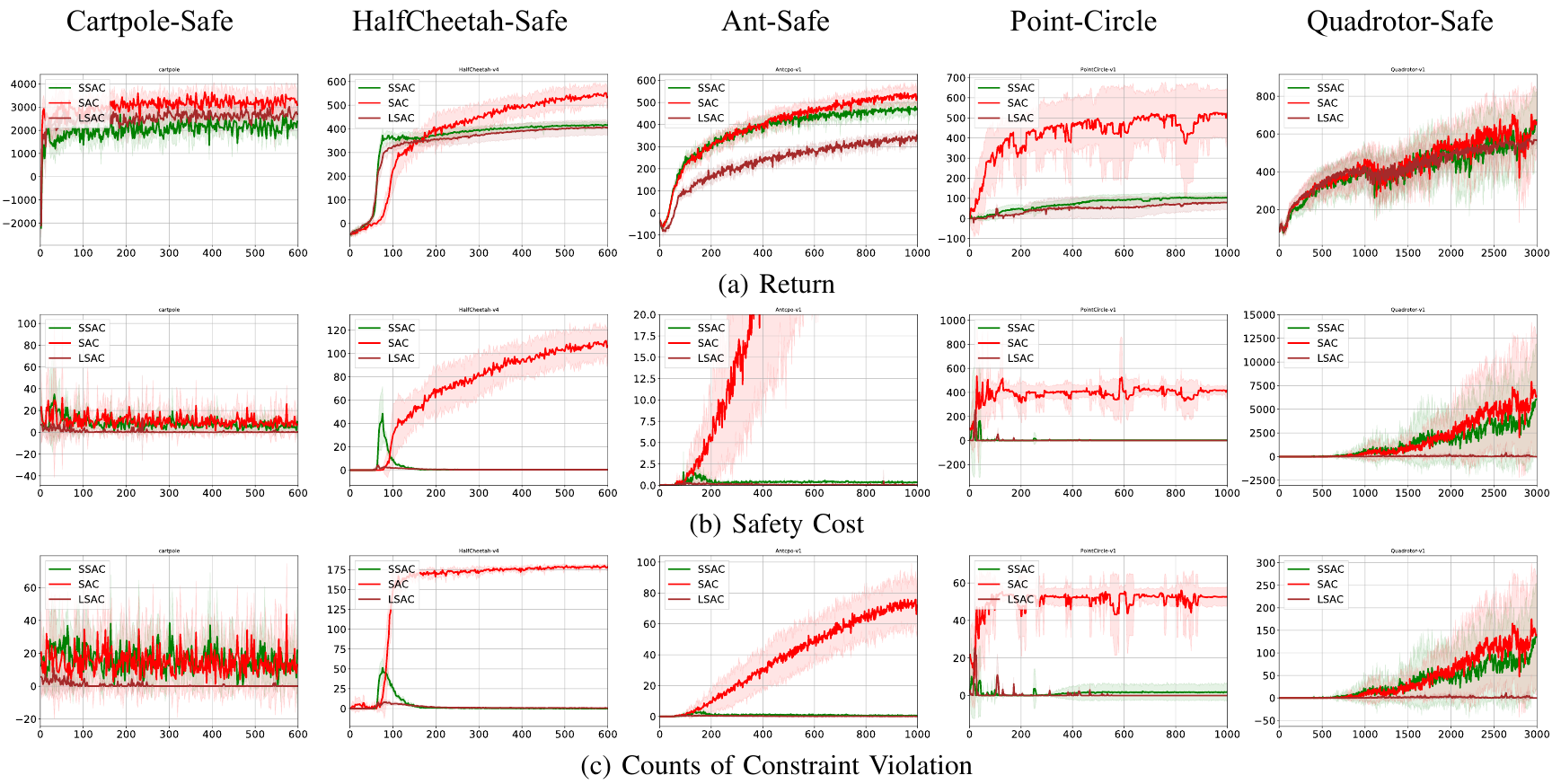}
\caption{The average performance of SAC (brown), LSAC (red), SSAC where the shaded areas show the 1-SD confidence intervals over 10 random seeds. The X-axis indicates the total time steps in thousands. Return and the sum of constraint functions are shown in the first and second row respectively.}
\label{fig2}
\end{figure*}

\subsection{Experiment Setups} \label{subsec:experiment setups}
In the following, five CMDP tasks are set up ranging from simulated robot locomotion in the MuJoCo simulator \cite{todorov2012mujoco} to motion planning of a simulated quadrotor: (i) Cartpole-Safe: The agent is rewarded for sustaining the pole vertically at a target position, while limited in a safe region; (ii \& iii) HalfCheetah-Safe \cite{chow2019lyapunov} and Ant-safe: The agent is rewarded for running while the speed is limited for safety; (iv) Point-Circle\cite{achiam2017constrained}: The agent is rewarded for following a circular trajectory while limited to stay in a safe region $\vert x\vert< \overline{x}$; (v) Quadrotor-Safe: The agent is rewarded for tracking a spiral trajectory while constrained to stay under a certain altitude. The details of the control tasks are described in the following.

\subsubsection{CartPole-Safe}

In this experiment, the agent is expected to sustain the pole vertically at a target position $x=6$. This is a modified version of CartPole with continuous action space \cite{brockman2016openai}. The action is the horizontal force $F$ on the cart, $F\in[-20, 20]$. While tracking the target position, the safety constraint $x<4$ needs to be ensured. The state space of $x$ is [0, 10] and the episodes end if the cart moves outside this region. The agent is initialized randomly as $x\in[0,4]$. The reward function is given by $ r=20\times \text{sign}(1 -|x-6|)\times(1 -|x-6|)^2+\text{sign}(\frac{\pi /2  - |\theta|} {\pi / 2})\times(\frac{\pi / 2 - |\theta|} {\pi / 2})^2 $.
The constraint function is given by $c = \max(|x|-3.2, 0)^2/5$.
The episodes are set at a length of $250$. 

\textcolor{black}{It should be noted that the target position is outside the safe region. As a result, the optimal safe behavior is to sustain the pole vertically at the edge of the safe region. This setup is deliberate in order to test whether safety can outweigh the return under the control of the trained agents.}

\subsubsection{Point-Circle}
This task is borrowed from \cite{achiam2017constrained}, where the agent controls a mass point to run in a wide circle but limited to stay in a safe region. The
agent is initialized at $(0,0)$. The agent is rewarded for moving the mass point counter-clockwise along a circle of radius 15m, $ r = \frac{-y \times v_x + x \times v_y}{1 + |\sqrt{(x^2 + y^2) - 15}|}$.
We constrain the x-axis position to be less than $2.4$, and the constraint function is: $c = \max(|x| - 2.4,0)$.
The episodes are set at a length of 65.

\subsubsection{HalfCheetah-Safe}
HalfCheetah-Safe is taken from \cite{chow2019lyapunov}. The task is to control a HalfCheetah (a 2-legged simulated robot with 17 degrees of freedom) to run as fast as possible with maximum velocity limited for safety.
The reward function is $r = v-0.1 \times \Vert a\Vert^2$ where $v$ is the forward speed of the HalfCheetah and $a$ is the control input.  The constraint function is $c =  \max(|v| - 2.7,0)^2$. The episodes are set at a length of $200$.

\subsubsection{Ant-Safe}
In Ant-Safe, the agent controls an Ant (a quadrupedal simulated robot) to run as fast as possible while satisfying the safety constraint on forwarding speed, $v<2.7$. The reward function is $r = v-0.5 * \sum a^2 - cost_\text{contact}+ 1$ where $cost_\text{contact}$ equals to $-1$ if the robot touches the ground and $0$ otherwise. The constraint function $c = \max(|v| - 2.7,0)^2$. The episodes are set at a length of 200. Reward function and other settings are \textcolor{black}{the same with the default setting in the OpenAI Gym}. The episodes are set at a length of 200.

\subsubsection{Quadrotor-Safe}

This task is to control a drone to track a spiral trajectory in 3D space. In the meantime, the altitude is limited to be under a threshold to avoid hitting the ceiling.
The simulation environment is modified from an open-source Crazyflie simulator Matlab code\footnote{\url{https://github.com/yrlu/quadrotor}}
into OpenAI Gym's structure. The simulator has three parts, the CrazyFlies model parameters, the PD controller, and the dynamic equations. The control proceeds as follows: The quadrotor simulator outputs the next state of the quadrotor given the force, torques, and the current state. The controller maps the observation of the current state to the desired next step and the desired state equals the current state adds the desired step. Last, the PD controller converts the current state and desired state to force and torques, and the loop continues. The state includes the quadrotor's position, velocities, attitude ,angular velocities and reference trajectory, $s_t=[x,y,z,\dot x,\dot y,\dot z, p, q, r,\dot p,\dot q,\dot r,x_{target},y_{target},z_{target}]$. The policy outputs desired relative changes in position and velocity from current state, $a_t=[\Delta x,\Delta y, \Delta z, \Delta \dot x,\Delta \dot y,\Delta \dot z]$. And an episode ends when the current position is too far from the reference trajectory. For this task, the reward is $ r =  -\Vert d \Vert+1$ where $d$ is the distance to the reference trajectory. The constraint function is $c = 100*\max(|z| - 0.4,0)$.
For this experiment, the episodes are set at a length of 2000.

\subsection{Evaluation and Comparison Analysis}

In this part, the performance of LSAC on the CMDP tasks is evaluated and compared with the existing safe RL algorithm, safe soft actor-critic (SSAC) \cite{chow2019lyapunov} with optimized hyperparameters. SSAC is a safety constrained variant of the original algorithms through the Lagrange relaxation procedure \cite{bertsekas1997nonlinear}. The soft actor-critic (SAC) \cite{haarnoja2018soft} is also included to show that the optimal behaviors are generally unsafe in the CMDP setting. 
In the meantime, the performance of LCPO is compared with constrained policy optimization (LCPO) \cite{achiam2017constrained}, a state-of-the-art trust-region method for CMDP tasks.
Since the trust-region methods (LCPO, CPO) require large batch sizes and need more samples than the gradient-based methods to reach convergence, thus, these two classes of methods are compared separately.
\textcolor{black}{The sum of the constraint function of episodes (safety cost) and the counts of constraint violations} is used as the measure of safety and the return is used to evaluate performance. The goal is to suppress these measures to zero while maximizing the return, i.e., the expectation of constraint function at each time step and the expectation of constraint violations are required to be zero.

\begin{figure*}[htb]
\centering
\includegraphics[width=1.0\textwidth]{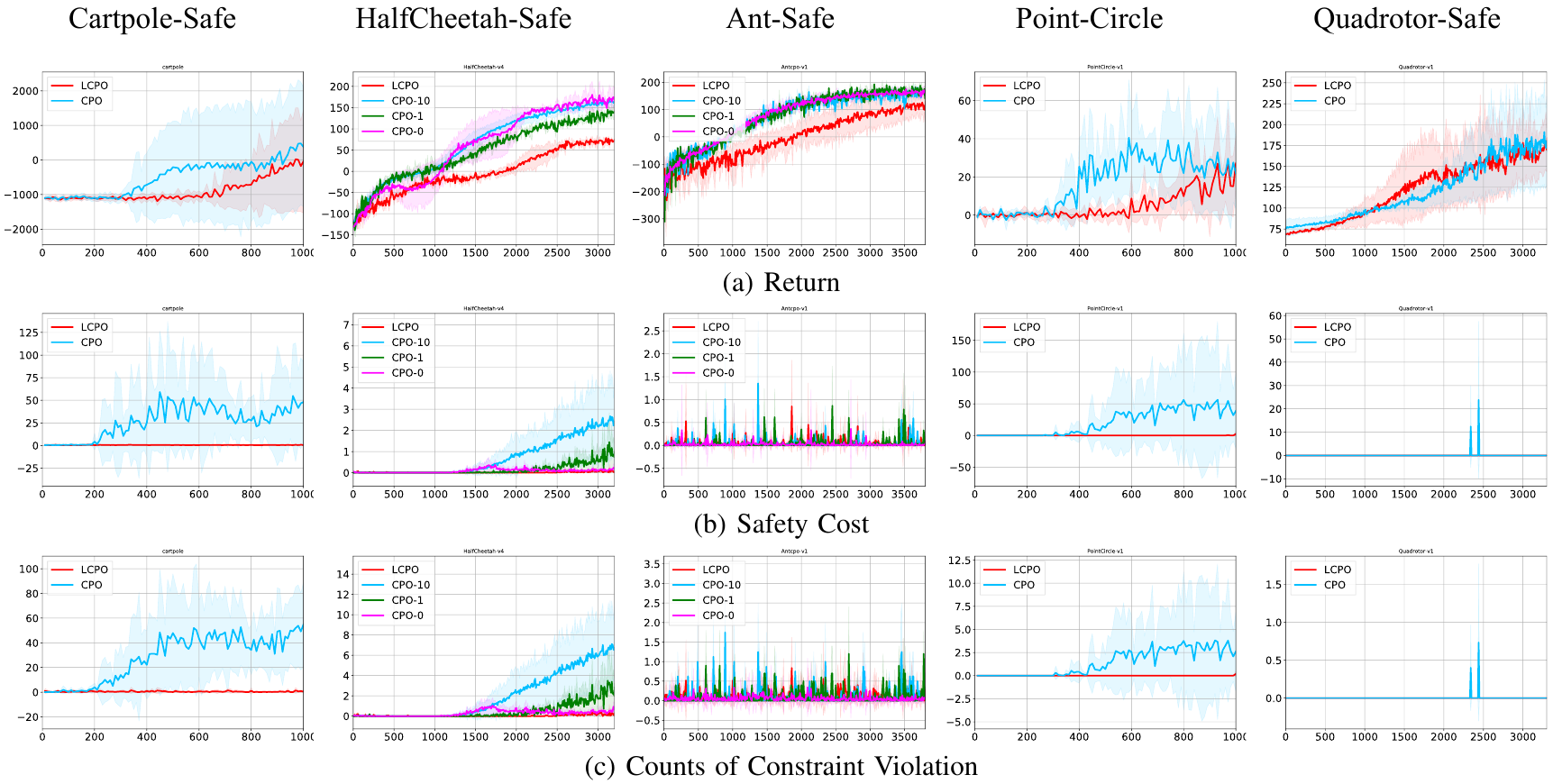}
\caption{\textcolor{black}{The average performance of CPO and LCPO, where the shaded areas show the 1-SD confidence intervals over 5 random seeds. The X-axis indicates the total time steps in thousands. The return (a), safety cost (b), and counts on constraint violations (c) are shown in each row respectively.}}
\label{fig-full comparison with CPO}
\end{figure*}

\subsubsection{Algorithm hyperparameters}
The hyperparameters for the LSAC and LCPO can be found in the Appendix.

For LSAC, there are three networks: the policy network, the Q network, and the Lyapunov network. For the policy network, we use a fully-connected MLP with two hidden layers of 256 units, outputting the mean and standard deviations of a Gaussian distribution. For the Q network and the Lyapunov network, we use a fully-connected MLP with two hidden layers of 256 units, outputting the Q value and the Lyapunov value. All the hidden layers use Relu activation function and we adopt the same invertible squashing function technique as \cite{haarnoja2018soft} to the output layer of the policy network. The hyperparameters can found in Fig.~\ref{Table-1}.

For LCPO, there are three networks: the policy network, the value network, and the Lyapunov network. For the policy network, it has two hidden layers of sizes (64, 32) with $\tanh$ activation functions, outputting the mean and standard deviations of a Gaussian distribution. For the value network, it has two hidden layers of sizes (256, 128) with Relu activation functions, outputting the value. And for the Lyapunov network, it has two hidden layers of sizes (256, 256) with Relu activation functions, outputting the Lyapunov value. The hyperparameters can found in Fig.~\ref{Table-2}.

The implementation of all the algorithms is based on TensorFlow \cite{abadi2016tensorflow}.

\subsubsection{Comparison with SSAC}
As demonstrated in Fig.~\ref{fig2}, though SSAC can maintain state constraint satisfaction on some of the tasks despite some major violations during training (see HalfCheetah-Safe and Ant-Safe), on other tasks it fails to find safe policy both at convergence or during training. On the other hand, our method (LSAC) quickly converges to safe policies across all the tasks (Fig.~\ref{fig2} (a)-(e)) while maintaining reasonable return. Besides, LSAC maintains low safety costs (almost zero) and constraint violation times throughout the training with low variance in all the tasks, even though all the policies are randomly initialized.

\subsubsection{Comparison with CPO}

\textcolor{black}{CPO and LCPO are compared on the safe environments as used for the off-policy methods. As shown in Fig.~\ref{fig-full comparison with CPO}, LCPO performs stably in terms of both safety cost and constraint violation counts in all of the environments, approximately achieving zero constraint violation during training and at convergence. In terms of return, LCPO performs comparably with CPO on Point-Circle and Quadrotor-Safe, while in other tasks CPO performs slightly better than LCPO, however, at a cost that CPO violates the safety constraints a lot, as shown in Fig.~\ref{fig-full comparison with CPO} (c). In CartPole and Point-Circle, CPO is not safe both during training and at convergence, while in other experiments it may violate constraints during training.}
As discussed in \cite{garcia2015comprehensive}, the discounted-sum constrained methods suffer from the tricky tuning of the safety threshold to handle the state constrained problems. To show this, we tested different threshold settings for CPO on Halfcheetah and Ant. In the HalfCheetah-Safe task, CPO swings due to the different settings of the safety threshold.
CPO either fails to find the feasible policy or reaches convergence after being unsafe for a long period (more than 500,000 steps). One more thing to note is that CPO requires additional cost shaping, by having a network evaluating the chance of constraint violation to achieve the best performance, while our approach doesn't need such techniques.

\subsection{Recoverability to Inner Set}
\begin{figure}[htb]
\centering
\subfigure[CartPole]{
\includegraphics[width=0.46\columnwidth]{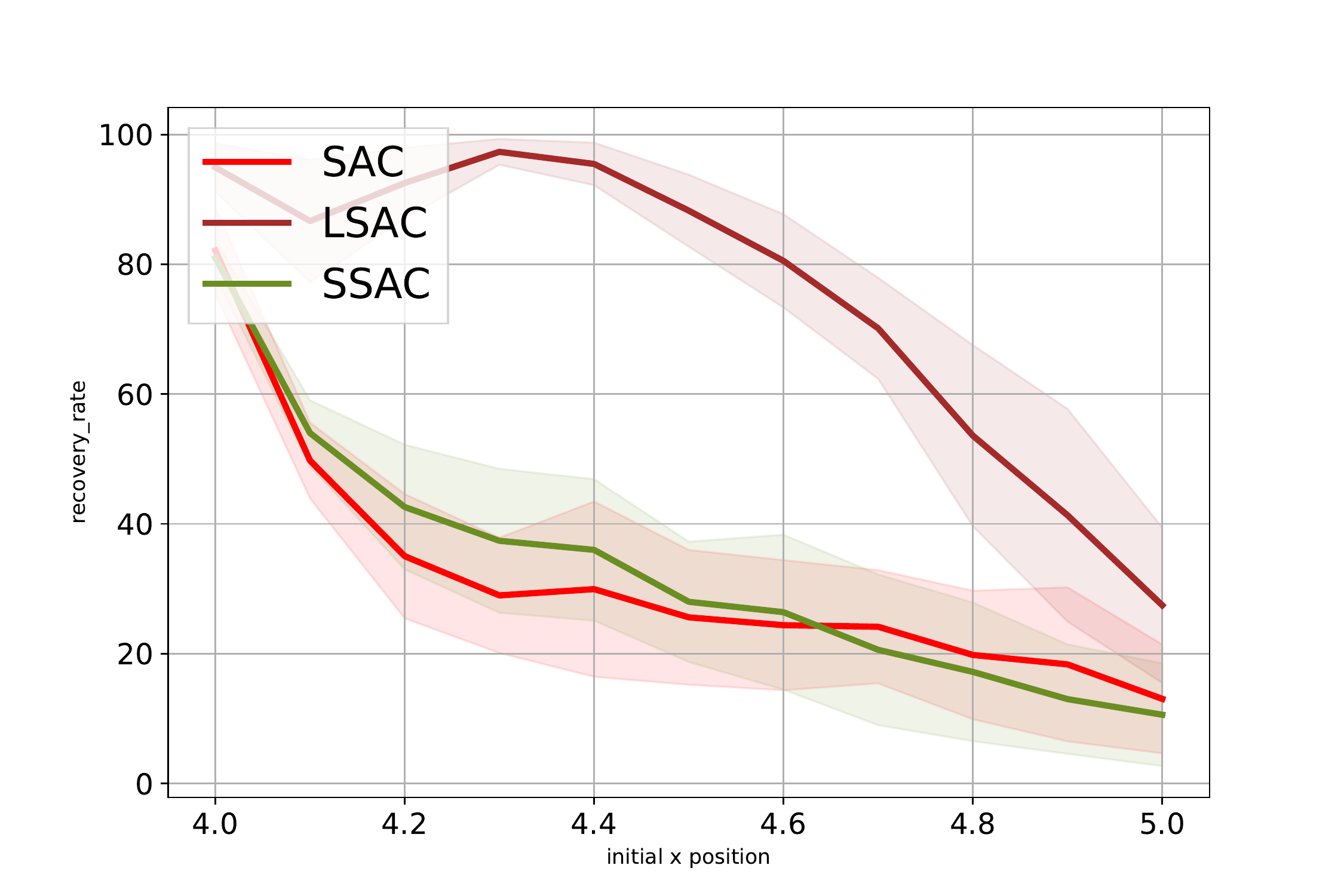}
}
\subfigure[Point-Circle]{
\includegraphics[width=0.46\columnwidth]{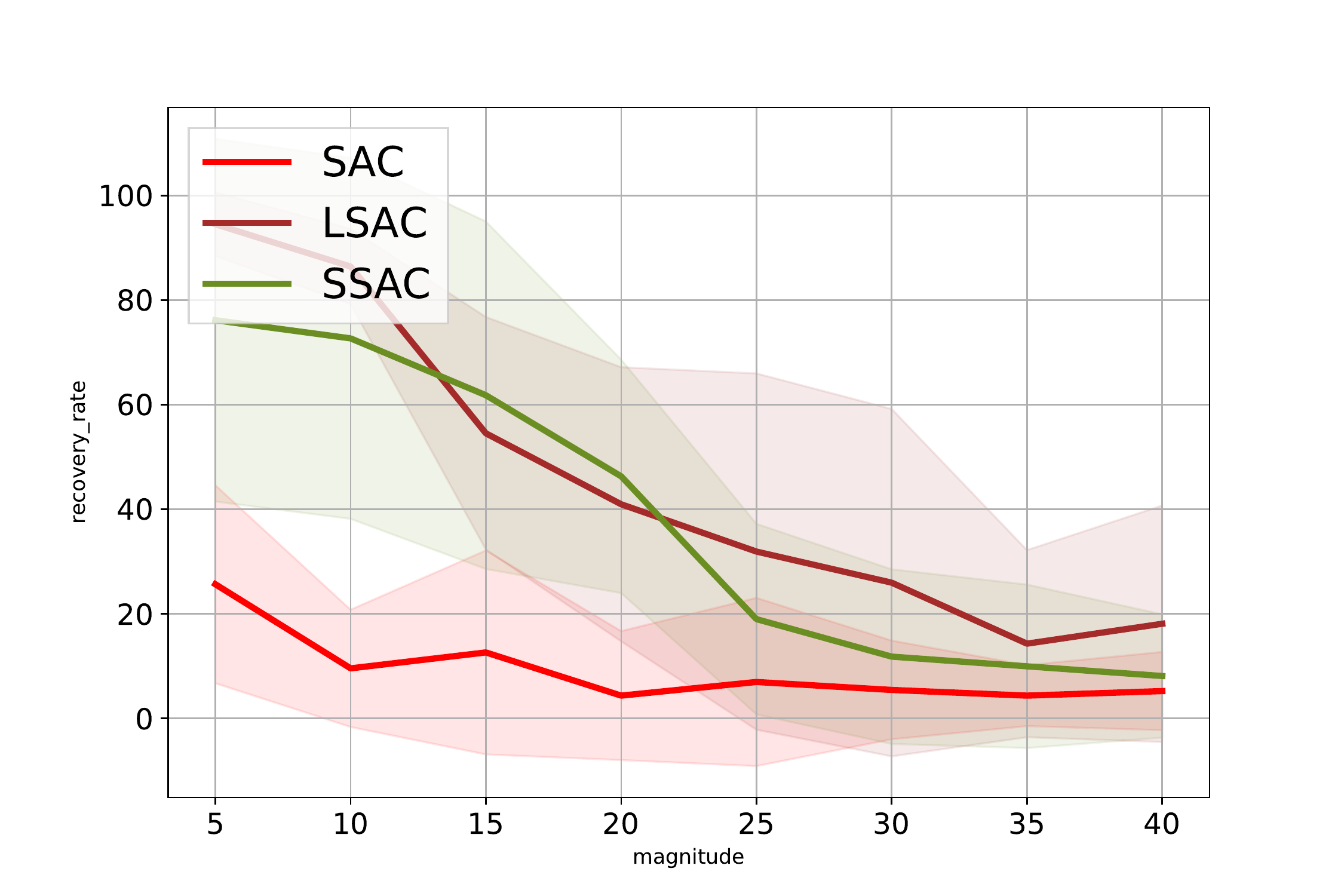}
}
\caption{The recovery rate of agents trained by LSAC, SSAC, and SAC in the presence of impulsive force $F$ with different magnitudes in Cartpole-Safe (a) and Point-Circle (b). The x-axis denotes the magnitude of an instant force. The policies with 10 different initialization are evaluated equally for 500 episodes under each magnitude of a force. The line indicates the average recovery rate of these policies and the shadowed region shows the 1-SD confidence interval.}
\label{fig1}
\end{figure}

As shown in Fig.~\ref{fig:state space} and Fig.~\ref{fig:timeview}, UUB assures that the system can recover to the inner set when it is wrongly initialized or accidentally disturbed and appears in the edge set. Now we evaluate the recoverability of the agents trained by LSAC and baselines when interfered by an unseen exogenous disturbance in CartPole-Safe and Point-Circle. Impulsive forces are implemented on the robot to push it outside the inner set and see whether it can recover to the inner set.
Their performance is measured by the recovery rate, i.e., the probability of successful recovery after impulsive disturbances. Under different impulse magnitudes, the policies trained by LSAC and baselines are evaluated 500 times, and the results are shown in Fig.~\ref{fig1}.

As observed in the figure, LSAC achieves the best performance in terms of recoverability when interfered by forces with different magnitudes, while SSAC is less possible to recover under the same circumstances. Note that the agent can not recover from arbitrarily large disturbances since only local UUB is assured.

\begin{figure}[htb]
\centering
\subfigure[Return]{
\includegraphics[width=0.46\columnwidth]{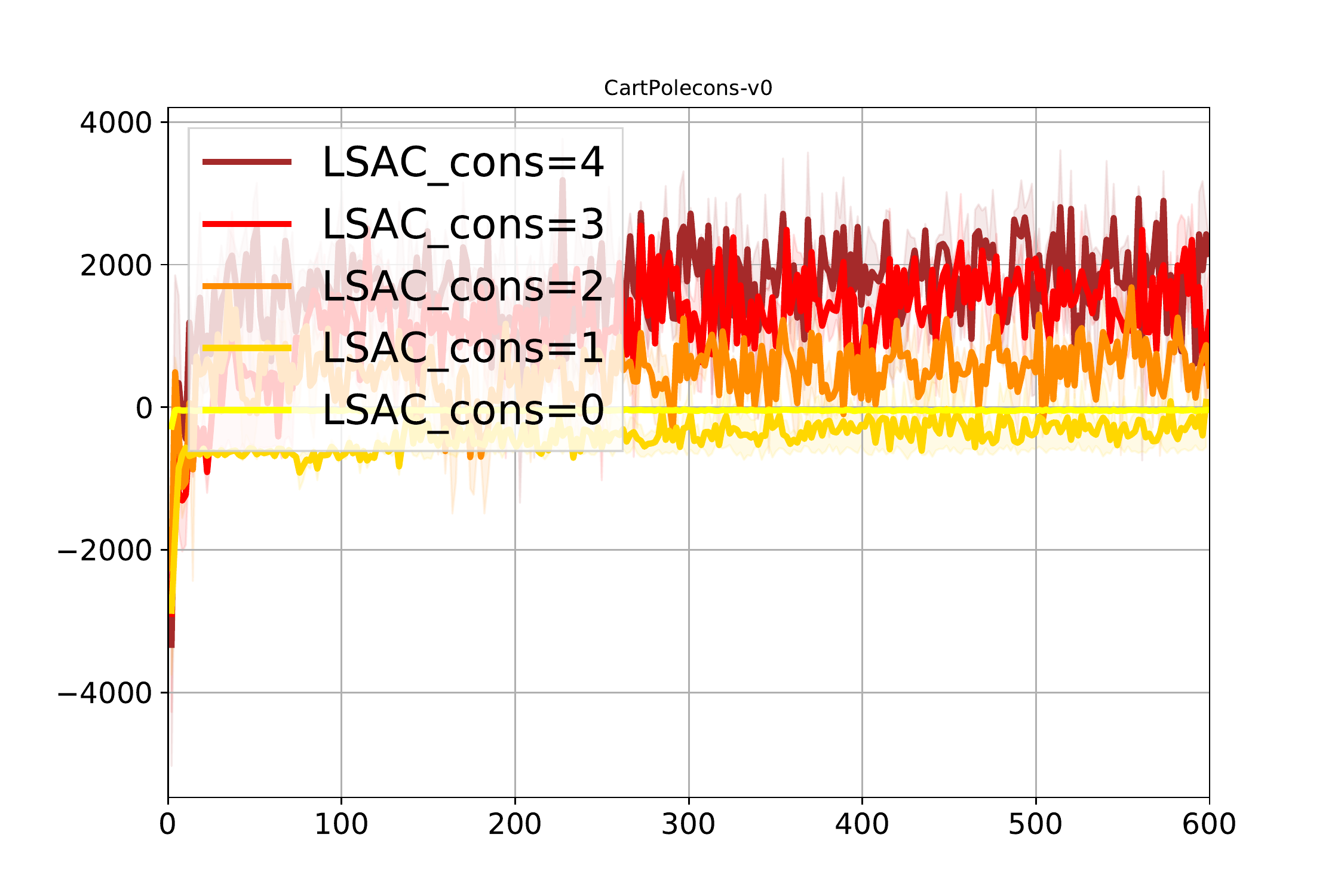}
}
\subfigure[Safety Cost]{
\includegraphics[width=0.46\columnwidth]{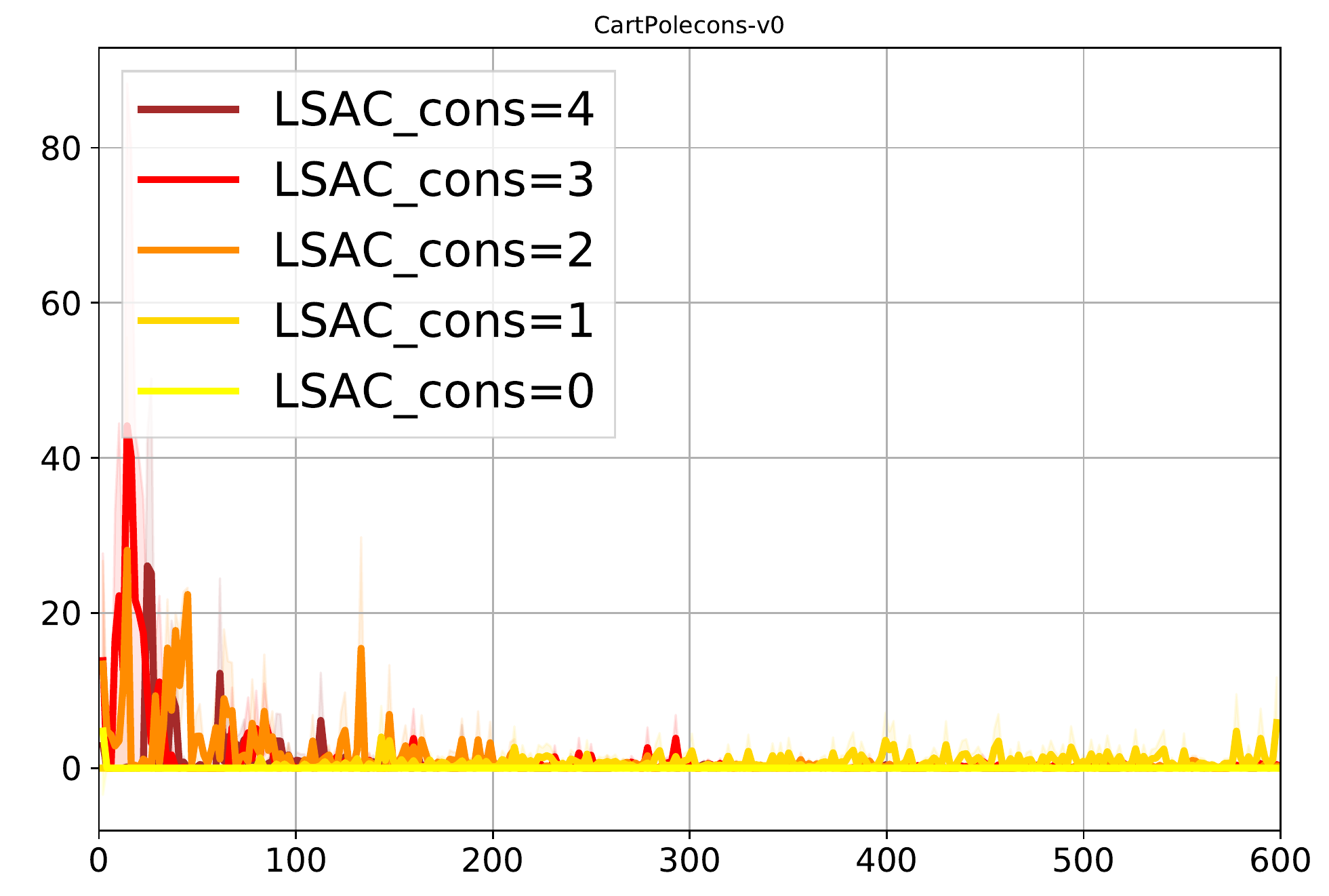}%
}
\caption{The average return and constraint function of LSAC in Cartpole-Safe with different constraints $\overline{x}$.}
\label{fig4}
\end{figure}

\subsection{Ablation on Constraints}
We want to test how does the proposed algorithm trade-off between performance and safety. In CartPole, the safety constraint is contradictory to the performance, i.e. being safer will decrease the return. Thus, we gradually strengthen the constraint and see how does LSAC reacts and when does it fail to find a feasible policy. Specifically, the size of inner set $\{x|x\in[0,\overline{x}]\}$ is reduced by assigning $\overline{x}$ with $\{0,1,2,3,4\}$.
The performance of LSAC in CartPole-Safe with different sizes of the inner set is compared, see Fig.~\ref{fig4}. 
As $\overline{x}$ approaches zero, the average return of LSAC also decreases while safety is maintained. However, when $\overline{x}=0$ and only the origin is safe, the agent fails to sustain the pole and dies almost immediately. This implies that LSAC may fail in the case that safety constraints are too strong.

\section{Conclusion}\label{sec:conclusion}

In this paper, a novel data-based approach for analyzing the uniformly ultimate bounded stability of a learning control system is proposed. Based on the theoretical results, two model-free reinforcement learning algorithms are developed, i.e., Lyapunov safe actor-critic and Lyapunov constrained policy optimization.
The proposed algorithms are evaluated on a series of robotic continuous control tasks with safety constraints. In comparison with the existing RL algorithms, the proposed method can \textcolor{black}{ reliably assure safety in various challenging continuous control tasks}. As a qualitative evaluation of stability, our method shows impressive resilience even in the presence of external disturbances.

For future work, we would like to explore the following directions: i) automating the design of constraint function; ii) extending the Lyapunov-based approach to model-based setting; iii) optimizing the performance upper bound while assuring the stability guarantee.

{\small
%\begin{table*}[htb]
\begin{figure*}
\centering
\caption{LSAC Hyperparameters}
\label{Table-1}
% \begin{center}
\begin{tabular}{l|c c c c c}\hline
Hyperparameters&Point-Circle&Ant&HalfCheetah&Quadrotor&CartPole-Safe\\
\hline
Lyapunov candidate function&Cost&Value&Value&Value&Value\\
Minibatch size& 256& 256& 256&256&256\\
Actor learning rate & 1e-4& 1e-4& 1e-4& 1e-4& 1e-4\\
Critic learning rate & 3e-4& 3e-4& 3e-4& 3e-4& 3e-4\\
Lyapunove learning rate & 3e-4& 3e-4& 3e-4& 3e-4& 3e-4\\
Target entropy&-2&-8&-6&-6& -1\\
Target smoothing coefficient($\tau$) &0.005&0.005&0.005&0.005&0.005\\
Discount($\gamma$) & 0.99 & 0.99 & 0.99 & 0.99 & 0.99  \\
$\alpha_3$&0.8 &1&1&0.8& 1 \\ 
\hline
\end{tabular}

% \end{center}
%\end{table*}
\end{figure*}
}

{\small
\begin{figure}[htb]
\centering
\caption{LCPO Hyperparameters}
\label{Table-2}
% \begin{center}
\begin{tabular}{l|c c c}\hline
Hyperparameters&Ant&HalfCheetah\\\hline
Lyapunov candidate function&Cost&Cost\\
Batch size& 10000& 10000\\
Critic learning rate & 1e-4& 1e-4\\
Lyapunov learning rate & 1e-4& 1e-4\\
Rollout length&500&500\\

Discount($\gamma$) & 0.99 & 0.99   \\
$\alpha_3$&0.2&0.2\\
maximum constraint value d&10&10\\
Stepsize &0.01 & 0.01\\
Safety discount &0.5&0.5 \\
\hline
\end{tabular}
% \end{center}
\end{figure}
}

%
%\bibliography{ref}
%\bibliographystyle{ieeetr}

\end{document}